\documentclass[10pt,prd,superscriptaddress,altaffilletter,nofootinbib,preprintnumbers,floatfix,notitlepage]{revtex4}

\usepackage{amsmath,amssymb,graphicx}
\usepackage{caption,subcaption} % for subfigures

\usepackage{color}

\usepackage[dvipsnames]{xcolor}

\begin{document}

\renewcommand{\arraystretch}{1.3} % makes the tables prettier
\setlength{\tabcolsep}{0.5em} % makes the tables prettier

\title{Cosmological dynamical systems in modified gravity}

\author{Christian G. B\"ohmer}
\email{c.boehmer@ucl.ac.uk}
\author{Erik Jensko}
\email{erik.jensko.19@ucl.ac.uk}
\affiliation{Department of Mathematics, University College London,
Gower Street, London WC1E 6BT, United Kingdom}
\author{Ruth Lazkoz}
\email{ruth.lazkoz@ehu.es}
\affiliation{Department of Physics, Faculty of Science and Technology, University of the Basque Country,  P.O.~Box 644, 48080 Bilbao, Spain}

\date{\today}

\begin{abstract}
The field equations of modified gravity theories, when considering a homogeneous and isotropic cosmological model, always become autonomous differential equations. This relies on the fact that in such models all variables only depend on cosmological time, or another suitably chosen time parameter. Consequently, the field equations can always be cast into the form of a dynamical system, a successful approach to study such models. We propose a perspective that is applicable to many different modified gravity models and relies on the standard cosmological density parameters only, making our choice of variables model independent. The drawback of our approach is a more complicated constraint equation. We demonstrate our procedure studying various modified gravity models and show how much generic information can be extracted before a specific model is considered.
\end{abstract}

\maketitle

\section{Introduction}

\subsection{Cosmology}
Modern cosmology has seen remarkable advances in recent times, and with it, our understanding of the Universe and the gravitational interaction has improved~\cite{LIGOScientific:2016aoc,Planck:2018vyg}. Despite being the weakest of the fundamental forces, it is responsible for the formation and evolution of structures on the largest scales. Hence, cosmology provides a unique testing ground to study the gravitational interaction where it dominates over the other forces.

The best current description of gravity is given by the theory of General Relativity (GR). Einstein's theory has been vastly successful in its predictive and explanatory power \cite{Will:2018}, though it does face challenges in the dark sector (i.e.~relating to dark matter and dark energy). With this in mind, one is encouraged to consider small modifications of General Relativity that look to solve these dark sector problems. In particular, there is no evidence against the presence of additional degrees of freedom that would reproduce the features of dark matter or dark energy, yet cause no impact within the regimes where GR is successful. 

Throughout this work we assume the validity of the cosmological principle, namely, that on sufficiently large scales the Universe is homogeneous and isotropic. Consequently, we can model the Universe using the Friedmann-Lema{\^i}tre-Robertson-Walker (FLRW) line element. Additionally, we restrict our study to spatially flat models, in agreement with most current observational data~\cite{Efstathiou:2020wem}. This line element reads
\begin{align}
    ds^2=-N^2(t)dt^2+a^2(t)\left[dx^2+dy^2+dz^2\right]\,,
\end{align}
where $a(t)$ is the scale factor and $N(t)$ is the lapse function. In some theories of modified gravity, it is not necessarily true that the lapse function can be set to one, which is why we keep it arbitrary for now. In fully diffeomorphism invariant theories one can always rescale the time coordinate to absorb the lapse function into the time coordinate.

It may be tempting to go beyond the scope of this work and study anisotropic cosmological models like the various Bianchi type models, as these offer interesting theoretical possibilities and contain a much richer structure that the FLRW model. However, anisotropic models are very tightly constrained~\cite{Saadeh:2016sak,Planck:2018vyg} and it is fair to say that there is no observational evidence that challenges the cosmological principle meaningfully.

FLRW cosmology, despite its mathematical simplicity, provides an excellent test bed for modified theories of gravity, see for instance~\cite{Copeland:2006wr,Jain:2007yk,Lombriser:2016yzn,Koyama:2015vza,Nunes:2016qyp,Koyama:2018som,Ishak:2018his,Lombriser:2018guo,Lazkoz:2019sjl,Benetti:2020hxp,Braglia:2020auw,DiValentino:2021izs} and references therein. Within the FLRW framework, competing gravitational theories give rise to differing cosmological predictions. Of those predictions, those relating to the current observational tensions are particularly relevant: specifically, the discrepancies around the velocity of receding astronomical bodies (as encoded in the $H_0$ value), and the present-day standard deviation of linear matter fluctuations on the scale of 8h$^{-1}$ (as best enciphered in the $S_8$ parameter that supersedes the well known $\sigma_8$ quantity). However, the current proposals of modified theories that aim to resolve these tensions, either individually or simultaneously, typically are imperfect in the sense they do not reconcile all cosmological parameters. This is precisely the motivation to continue exploring modified gravity models from as many complementary angles as possible.

When considering a homogeneous and isotropic cosmological background, one can take advantage of the ability to cast the field equations of modified gravity theories into those of dynamical systems. It is then possible to explore the evolutionary consequences of the additional degrees of freedom. To this end, we will specify matter-energy content consisting of a matter and a radiation component along with a cosmological constant. This offers the possibility to compare our new scenarios with the consensus $\Lambda$CDM model, which is characterised by exhibiting a de Sitter point (cosmological constant dominated) in the asymptotic final state of its evolution, complemented with a radiation dominated repeller (instability along all eigendirections) and a matter dominated saddle point (instability along some eigendirections only). We present an approach of great generality in the definition of variables that will allow us to discuss in detail the differences from the $\Lambda$CDM model phase space pattern, and it may serve in the future as a criterion to discard largely discrepant models. 

\subsection{Modified gravity}

Theories of gravity beyond General Relativity are almost as old as GR itself. Given that GR is formulated on a four dimensional Lorentzian manifold that is torsion free and comes equipped with a metric compatible connection, one is immediately tempted to allow for additional geometrical structures or to increase the number of dimensions. The Einstein-Hilbert action is linear in the curvature scalar, which motivates models with nonlinearities. Matter is coupled minimally in GR, which means it might also be of interest to consider non-minimal couplings. The vast majority of modified gravity models being considered fall somewhere in the above description~\cite{Capozziello:2002rd,Ferraro:2006jd,Sotiriou:2008rp,DeFelice:2010aj,Nojiri:2010wj,Capozziello:2011et,Harko:2011kv,Clifton:2011jh,Bamba:2012cp,Nesseris:2013jea,Joyce:2014kja,Cai:2015emx,Nojiri:2017ncd,Boehmer:2021aji,CANTATA:2021ktz,Bohmer:2021sjf}.

For the purpose of the present work, we are particularly interested in second order modified theories of gravity. By this we mean that the gravitational field equations contain at most second derivatives with respect to the independent variables. Those independent variables are the four coordinates, one temporal coordinate, and three spatial coordinates. Given that the overall scale of the Universe should not enter the dynamical equations, other than through a cosmological constant, the cosmological equations of any such modified theory of gravity must take the form
\begin{align} \label{E1}
    E_1(H,\dot{H}) &= 8\pi \kappa\, \rho_{\rm matter} + 8\pi \kappa\, \rho_{\rm other}\,,\\
    E_2(H,\dot{H}) &= 8\pi \kappa\, p_{\rm matter} + 8\pi \kappa\, p_{\rm other}\,.
\end{align}
Here $E_1$ and $E_2$ are two arbitrary functions. This framework is very general and can be used to describe a plethora of second order modified gravity theories. In fact, without introducing non-minimal couplings, all such theories are probably of that form. Next, one divides these two equations by $H^2$ and introduces the standard density parameters on the right-hand sides to get
\begin{align}
    \tilde{E}_1(H,\dot{H}) &= \Omega_{\rm matter} + \Omega_{\rm other}\,,
    \label{E2a} \\
    \tilde{E}_2(H,\dot{H}) &= w_{\rm matter} \Omega_{\rm matter} + w_{\rm other} \Omega_{\rm other}\,, 
    \label{E2b}
\end{align}
where we have assumed linear equations of state. This approach can also include scalar fields, in which case the effective equation of state parameter $w$ would not be a constant. If, at this point, the matter satisfies the usual conservation equation (again neglecting non-minimal couplings), a first order equation in the matter and geometrical variables, it is immediately clear that one deals with a first order system of equations. We continue to elaborate on this point in the following subsection.

\subsection{Dynamical systems}

A large body of literature already exists which discusses dynamical systems applications in cosmology, see~\cite{Bahamonde:2017ize} and the many references therein. In the following, we will mainly discuss the choice of variables and the different approaches taken in the past, as this is one of the most relevant topics in the context of the current work. 
The starting point of the dynamical systems approach has often been the Friedmann constraint equation, which we write in the form
\begin{align}
    3 f^{(1)} H^2 + f^{(2)} = 8\pi \kappa\, \rho_{\rm matter} + 8\pi \kappa\, \rho_{\rm other}\,,
    \label{int:dyn1}
\end{align}
where $f^{(1)}$ and $f^{(2)}$ are arbitrary functions depending on the scale factor and its derivatives and $\rho_{\rm other}$ stands for `matter' terms other than the standard dust or radiation. The cosmological field equations in $f(R)$ gravity, $f(T)$ gravity, $f(Q)$ gravity and $f(\mathbf{G})$ gravity can all be put into this form. In fact, it is difficult to envisage a model which could not be brought into this form. We note that this is slightly less generic than the above mentioned equations~(\ref{E2a}) and~(\ref{E2b}). Equation~(\ref{int:dyn1}) reduces to General Relativity when $f^{(1)}=1$ and $f^{(2)}=0$. 

When working with dynamical systems it is most convenient to introduce dimensionless quantities. This has motivated the following approach when dealing with the Friedmann constraint~(\ref{int:dyn1}). The dimensions of $H^2$ and $\kappa \rho_{\rm matter}$ are the same, hence $f^{(1)}$ is dimensionless while $f^{(2)}$ also has dimensions of $H^2$. Dimensionless variables now appear naturally when equation~(\ref{int:dyn1}) is divided by a quantity that has the dimensions of $H^2$. In standard cosmology with scalar fields, see for instance~\cite{Copeland:1997et}, it is natural to simply divide by $H^2$ which leads to the cosmological density parameters and the constraint equation takes a particularly simple form. In $f(R)$ gravity, on the other hand, it appears to be natural to divide by the factor $3f^{(1)}H^2$ and to introduce dynamical variables that also depend on the chosen function~\cite{Amendola:2006we}. This leads to a neat set of equations, but can also introduce problems when $f^{(1)}$ vanishes dynamically during the cosmological evolution, the constraint equation again takes a rather simple form. The mentioned problem has motivated the introduction of other variables which remain regular during the cosmological evolution, see in particular~\cite{Carloni:2015jla} and~\cite{Alho:2016gzi} where specific $f(R)$ gravity models were studied, see also~\cite{Carloni:2018yoz,Rosa:2019ejh,Chakraborty:2021mcf}. Often, these improved variables lead to more complicated constraint equations.

Our approach follows a slightly different route. Instead of looking for particularly chosen variables applicable to certain modified gravity models, we simply divide our Friedmann constraint equation by $H^2$ and introduce the standard cosmological density parameters. The price to pay for keeping this simplicity is an involved constraint equation. The substantial gain from this is that we can apply this method independently of the functions that determine the model. In fact, we can study any model where the function depends on the Hubble function. Moreover, we can present a partial analysis of the model for all functions and extract the generic dynamics for a large class of models. The existence and location of all critical points and critical/singular lines only depends on the values of the function and its first and second derivatives at a finite set of points. Often, we can determine the stability properties of the critical points without having to specify the model altogether. This happens when the eigenvalues give definite values which are independent of the model.

Irrespective of the specific function chosen, we find that the de Sitter point (when it exists) is a stable late time attractor. This is a surprisingly general result which holds in many different modified theories of gravity. It shows that a late time accelerating cosmological solution is a truly generic property of the cosmological models we study.

The principal ideas of our approach are similar to those used in~\cite{Hohmann:2017jao}, applied to $f(T)$ gravity, and predating some of the more recent work on other second order modified gravity theories where their approach can also be used. The authors introduce what they call the Friedmann function and rewrite the dynamical equations using this function and its derivative. Their approach uses a dimensionless variable related to the ratio of matter and radiation. However, to recover the standard cosmological matter and radiation densities, the Friedmann function is needed, meaning that it is somewhat more difficult to obtain model independent statements about the matter content at certain critical points.

\subsection{Modified gravity models -- $f(\mathbf{G})$, $f(T)$ and $f(Q)$}
\label{section:fqmodels}

In the following, we deal with three classes of modified gravity theories which are generally known as $f(\mathbf{G})$, $f(T)$ and $f(Q)$ gravity, all of which can be found in~\cite{BeltranJimenez:2017tkd,Harko:2018gxr,BeltranJimenez:2019tme,Boehmer:2021aji,CANTATA:2021ktz} and the previously mentioned overview papers. All these models have in common that the cosmological field equations are compatible with the choice $N=1$ for the lapse function. Moreover, the cosmological equations are also of second order only. All models are based on a single scalar (or pseudo-scalar in the case of $f(\mathbf{G})$ gravity), which is indirectly linked to the standard curvature scalar.

When considering spatially flat FLRW models, it turns out that the relevant scalars defining the respective theories are given by
\begin{align}
    \mathbf{G} = -6\frac{H^2}{N^2}\,, \quad 
    T = 6\frac{H^2}{N^2}\,, \quad 
    Q = 6\frac{H^2}{N^2}\,.
\label{scalars}
\end{align}
Perhaps unsurprisingly, the cosmological field equations implied by these different models all coincide provided the signs are taken care of and some minor re-definitions are made to compensate for unimportant factors.
 
 For all three classes of modified gravity models the cosmological field equations can be written as
\begin{align}
    6 F H^2-\frac{1}{2}f &= \rho \,, 
    \label{Friedmann}\\
    (12 H^2 F'+F)\dot H &= -\frac{1}{2}(\rho+p) 
    \label{Ray} \,.
\end{align}
The reason for this equivalence can be understood by carefully studying the various terms which connect these theories which are all defined in different geometrical settings. The relevant terms were identified in~\cite{Boehmer:2021aji} where equivalence for more general spacetimes was also discussed. For all three models, the energy-momentum conservation equation is automatically satisfied and does not yield additional conditions to be satisfied. This is due to either the choice of suitable tetrads or coordinates, depending on the model in question.

A cosmological constant $\Lambda$ can always be included in these models by using $f \to f + 2\Lambda$. Here $f$ is the function specifying the model and
\begin{align}
    F = \frac{df}{d\mathbf{G}} \quad \text{or} \quad
    F = \frac{df}{dT} \quad \text{or} \quad
    F = \frac{df}{dQ} \,,
\end{align}
depending on the geometric setting. Likewise, $F'$ stands for the second derivative with respect to the argument. Note that the field equations of $f(R)$ gravity, for example, cannot be brought into this form as this is a fourth order theory while all the above are second order theories.

The following discussion will be valid for all of the three mentioned models. However, in order not to repeat this fact many times and for ease of notation, we will sometimes use $Q=6H^2$ and will often work with $Q$. All of the following could be formulated equally using either $T$ or $\mathbf{G}$.

\section{Dynamical system -- standard approach}

\subsection{Towards a dynamical systems formulation}
 
The identification of suitable variables for a dynamical systems formulation is generally challenging for modified gravity models, see~\cite{Bahamonde:2017ize}. One can often introduce distinct, well-motivated variables which lead to distinct dynamical systems with different features~\cite{Amendola:2006we,Carloni:2015jla,Alho:2016gzi}. Following one of the standard approaches, we begin with the Friedmann equation~(\ref{Friedmann}) and divide by $6F H^2$ which gives
\begin{align}
    1 = \frac{\rho}{6F H^2} + \frac{f}{12F H^2} \,,
\end{align}
 where we assume $F \neq 0$. This only excludes constant functions which are of no interest to us. Since the signs of both $f$ and $F$ are undetermined, in general, we begin by introducing the two variables
 \begin{align}
    X = \frac{\rho}{6F H^2}\,, \qquad Y = \frac{f}{12F H^2} \,,
    \qquad X + Y = 1\,. 
    \label{XY1}
\end{align}

Should the model under consideration contain multiple matter sources like matter and radiation, one would naturally introduce several different variables $X_i$, one for each matter source.

The two variables $X$ and $Y$ are not sufficient to rewrite the field equations in the form of a dynamical system as one is not able to eliminate the Hubble function from the equations. We address this issue by introducing a third variable, related to the Hubble function. As $Q$ is related to the Hubble function via $Q=6H^2$, we have $Q \geq 0$ at all times. The new variable is
\begin{align}
    z = \frac{Q}{Q_0+Q} = \frac{H^2}{H_0^2+H^2}\,,
    \label{eqn:zq}
\end{align}
and satisfies $0\leq z \leq 1$ for all times, which corresponds to $0\leq Q < \infty$. Clearly one can always introduce such a variable irrespective of the modified gravity model that is being considered. Here $H_0$ is a constant Hubble parameter and $Q_0=6H_0^2$. One would then expect to have two independent variables, either $\{X,z\}$ or $\{Y,z\}$, for one matter source or $N+1$ independent variables for $N$ matter (fluid with given equation of state) sources. 

However, as any function of $H$ can be rewritten in terms of $z$, the variable $Y$ is not strictly independent. We shall see in the following section that the system can be reduced by eliminating $Y$ altogether, despite its apparent necessity in (\ref{XY1}). Here we shall allow $Y$ to represent the free parameters that may occur in $f$ (e.g.~a cosmological constant term) but are not captured in $z$. The Friedmann constraint (\ref{XY1}) then determines the values of the free parameters at each point of the phase space. If there are no free parameters, then (\ref{XY1}) singles out the allowed 1D trajectory. This will be elucidated in the concrete examples below.

The introduction of a scalar field would require the introduction of two further variables, one related to its kinetic energy and one for the potential. The dynamical equations for $X$ and $z$ are given by
\begin{align}
    \frac{dX}{dN} &= 3X\left( (X-1) + \frac{X}{1+m(z)}\right) := f = X \tilde{f}\,,
    \label{eqn:gen1a}\\
    \frac{dz}{dN} &= -6X\frac{(1-z)z}{1+m(z)} := g = X \tilde{g}\,,
    \label{eqn:gen1b}
\end{align}
where $N=\log a$. The functions $f,g,\tilde{f},\tilde{g}$ are defined through the above equations. The function $m(z)$ is defined by
\begin{align}
    m = \frac{12H^2 F'}{F}\,,
    \label{eqn:gen2}
\end{align}
taking into account the relation between $z$ and $H$ as given in~(\ref{eqn:zq}). One can now make various statement about this system without specifying the function $f$ or equivalently the function $m(z)$. Let us remark that the above system is not well-defined if $f=c_1 \sqrt{6H^2} + c_2$ as this gives $m=-1$. In this case, the original equations should be considered separately for this particular choice. We will not discuss this particular case separately as solutions of this type were already addressed in~\cite{BeltranJimenez:2017tkd}, in the context of $f(Q)$ gravity.

The line $X=0$ needs to be treated carefully as both dynamical equations vanish along this line. The location of all possible critical points is established as follows. Excluding $X=0$ the first equation can vanish when $X=(1+m(z))/(2+m(z))$, assuming $2+m(z)\neq 0$. This means, any critical point will have this $X$-coordinate for any value of $z$ that makes the right-hand side of the $z$-equation vanish. Clearly, we have the two values $z=0$ and $z=1$ provided that $2+m(0)\neq 0$ and $2+m(1)\neq 0$. Moreover, any value $z_c$ for which $m(z_c)=-1$ corresponds to a singular line as the system would be ill-defined. On the other hand, if $z_\star$ is such that $m(z_\star) \rightarrow \infty$, we would also find critical points. Let us summarise these general finding in Table~\ref{tab1} as follows:

\begin{table}[htb!]
    \centering
    \begin{tabular}{c|c|c|p{5cm}}
         Point/Line & $X$ & $z$ & condition \\ \hline\hline
         L1 & $X=0$ & any & critical line for all $z$ \\
         L2 & any & $z_c$ & any $z_c$ for which $m(z_c)=-1$ exists \\
         A & $X=\displaystyle\frac{1+m(0)}{2+m(0)}$ & $z=0$ & $2+m(0) \neq 0$ \\[8pt]
         B & $X=\displaystyle\frac{1+m(1)}{2+m(1)}$ & $z=1$ & $2+m(1) \neq 0$ \\
         C & $X=1$ & $z=z_\star$ & $m(z_\star) \rightarrow \infty$ exits \\
    \end{tabular}
    \caption{Location of critical points for generic model.}
    \label{tab1}
\end{table}
Let us note that for some particular functions, things could be rather contrived. For instance, a function for which $m(1) \rightarrow \infty$, would mean that B and C would not exist independently. Other particular scenarios could be set up. When discussing specific models, we highlight these issues explicitly. Lastly, while the range of the variable $z$ is bounded from below and above, things are less clear for $X$. If $F$ cannot change sign, we would have $0\leq X$, however, there would be no obvious upper bound. This means that the generic phase space is a strip of height one and possibly infinite width. Certain models will reduce the size of this strip. We will later use a \emph{poor mans version} of compactification where we introduce $\tilde{X}=X/(1+X)$ which is bounded from above by $1$ as our $X$ will be positive, see also~\cite{Hohmann:2017jao}.  

Let us also note for future reference that the $X$-nullclines are given by the curve $X=(1+m(z))/(2+m(z))$ while the $z$-nullclines are located along the lines $z=0$ and $z=1$. 

\subsection{Stability discussion}

The critical points of system~(\ref{eqn:gen1a})--(\ref{eqn:gen1b}), as given in Table~\ref{tab1}, depend on the explicit form of $m(z)$. Nonetheless it is possible to make quite general statements about the nature of the critical points. The stability matrix or Jacobian is given by
\begin{align} \label{Jacobian}
    J = \begin{pmatrix}
   \displaystyle \frac{\partial f}{\partial X} &\displaystyle \frac{\partial f}{\partial z} \\[7pt]
 \displaystyle   \frac{\partial g}{\partial X} &\displaystyle \frac{\partial g}{\partial z} 
    \end{pmatrix} \,.
\end{align}
Let us denote the two eigenvalues of $J$ by $\lambda_1$ and $\lambda_2$. A direct calculation shows that for $X=0$, which is the line L1, the eigenvalues are $(0,3)$ for every point on this line. Therefore, the system is not hyperbolic along L1, this is perhaps expected for a critical line. The trajectories or orbits of the system satisfy $dX/dz = f/g = \tilde{f}/\tilde{g}$, which means these are unaffected when $X \neq 0$. However, the direction of the trajectories with respect to time does change, they reverse for $X<0$ but do not change for $X>0$. This simple observation is sufficient to determine whether the line $X=0$ attracts or repels trajectories. The properties of the line L2 are more difficult to establish as they depend on $m'(z)$, the first derivative of $m$ with respect to $z$.

For point A, one directly finds $\lambda_1 = 3$ and $\lambda_2 = -6/(2+m(0))$. Hence, point A is never stable. Depending on the value of $m(0)$ this point is either a saddle point or a repeller. 

Point B is challenging to study as  $z=1$ corresponds to $6H^2\rightarrow\infty$ which can pose problems in the definition of $m$, see~(\ref{eqn:gen2}). One can show that $\lambda_1 = 3(1+m(1))/(2+m(1))$, however the second eigenvalue can diverge in the limit $z \rightarrow 1$ if $m'(1)$ does not decrease fast enough. Likewise, for point C one can establish $\lambda_1 = 3$ but the second eigenvalues will again depend on the derivative of $m$. This means that point C cannot be stable as one eigenvalue is always positive. Likewise, point B cannot be stable if $m(1)>-1$ or $m(1)<-2$. For the parameter range $-2<m(1)<-1$ the stability will depend on the other eigenvalue provided it is well defined. 

We will look at some examples which discuss the above features, almost all of which are absent when studying the simple case of General Relativity. 

\subsection{General Relativity}

It makes sense to briefly revisit the simple case of General Relativity, where $f(6H^2)=6H^2 + 2\lambda_0$ so that $F=1$ and $F'=0$ which implies $m=0$. Here $\lambda_0$ stands for a cosmological constant term. This means the line L2 does not exist and the point C also does not exist either. The line L1 is present and we have two critical points A and B. The variable $X$ now becomes $X={\rho}/{(6H^2)}=\Omega_{\rm m}/2$ where $\Omega_{\rm m}$ is the standard matter density parameter, hence $0 \leq X \leq 1/2$. Likewise, for $Y$ we have $Y=1/2 + \lambda_0/6/H^2$ or $Y = 1/2 + \Omega_{\lambda_0}/2$. Together with $0\leq z \leq 1$ this reduces the phase space to a rectangle. 

From the Friedmann constraint $X+Y=1$, one has that each $X$ value fixes $Y$. And as mentioned previously, the $Y$ variable can be rewritten in terms of $z$ plus free parameters. Therefore each point $\{X,z\}$ uniquely determines the value of the free parameters in $f$. In other words, each trajectory represents the evolution of $\rho_m$ and $H$ for a different value of $\lambda_0$. 

We can now interpret the two critical points. At point A when $z=0$ we have $H=0$, which immediately gives us a static solution. The finite value of $X=1/2$ implies that $\rho$ must vanish at the same rate as $H^2$ so that we have a static vacuum solution. This solution is only consistent with $\lambda_0=0$ and corresponds to Minkowski space. 

For point B we have $z=1$, which corresponds to $H \rightarrow \infty$. As $X=1/2$ is finite there we must also have $\rho \rightarrow \infty$. Therefore $1/2 = \rho/6/H^2 = \rho_0/a^3/6/H^2$. This is solved by $a(t)=a_0 t^{2/3}$ as one would expect for the matter dominated universe. 

\begin{figure}[htb]
    \centering
    \includegraphics[width=0.5\textwidth]{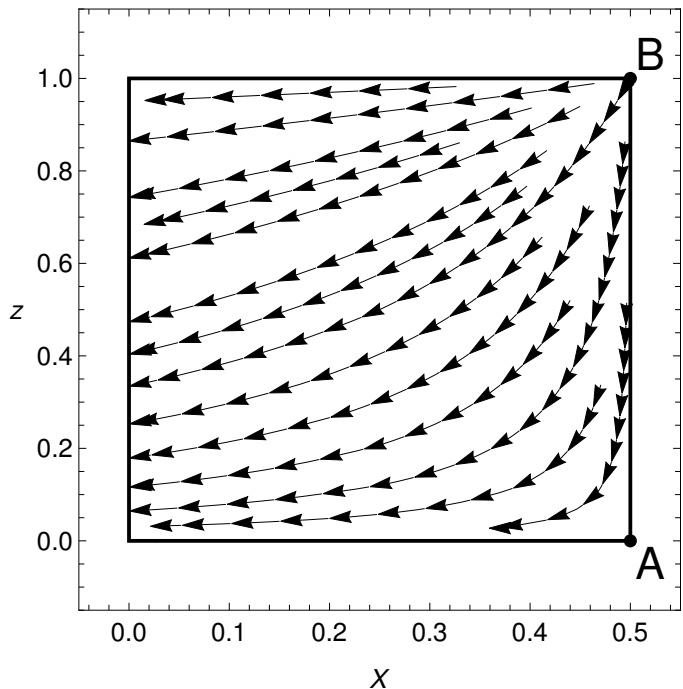}
    \caption{General Relativity phase space using the variables $X$ and $z$.}
    \label{fig:GR1}
 \end{figure}

This leads to the following conclusions: the matter-dominated universe is the early time repeller from which all trajectories start. All trajectories are then attracted towards the line $X=0$ where they terminate at some value $z$. If we denote one such value by $z_0$ then we will have a corresponding $H_0$. When $X=0$ we have $Y=1$ which means
\begin{align}
    1 = Y = \frac{6H_0^2 + 2\lambda_0}{12H_0^2} =
    \frac{1}{2} + \frac{\lambda_0}{6H_0^2}\,,
\end{align}
which gives %$\frac{\lambda_0}{3H_0^2}=1$ or 
$H_0 = \sqrt{\lambda_0/3}$. Consequently, the line $X=0$ corresponds to de Sitter type solutions parameterised by different values of $H_0$. In this formulation of General Relativity with a cosmological term we find an early time universe that is matter dominated, which then evolves towards a dark energy dominated universe. This is visualised in Fig.~\ref{fig:GR1}.

 One can state various explicit results due to the simplicity of the equations when $m=0$. One can immediately integrate the $X$-equation~(\ref{eqn:gen1a}) which gives
 \begin{align}
    X = \frac{1}{2+c_1 \exp(3N)} \,,
\end{align}
where $c_1$ is a constant of integration. In line with the above discussion we have $\lim_{N\rightarrow -\infty}X = 1/2$ and $\lim_{N\rightarrow \infty}X = 0$ confirming that all trajectories begin at $X=1/2$ and terminate at $X=0$. Dividing both equations~(\ref{eqn:gen1a}) and~(\ref{eqn:gen1b}) yields
\begin{align}
    \frac{dz}{dX} = -\frac{2(1-z)z}{(2X-1)} \,,
\end{align}
which can easily be solved using separation of variables. This leads to
\begin{align}
    z = \frac{1}{1+c_2(1-2X)} \,.
\end{align}
Consequently, all trajectories starting at $X=1/2$ satisfy $z=1$, as expected, which corresponds to point B. When $X=0$ we have $z=1/(1+c_2)$ which determines vertical position along the $X=0$ axis. As stated earlier, this fixes the Hubble constant and corresponds to a de Sitter type solution.

\subsection{Dimensional reduction}
\label{sec:dimred}

The previous discussion follows the `standard' approach of writing the cosmological field equations in the form of a dynamical system. This standard approach is based on the idea of rewriting the Friedmann equation~(\ref{Friedmann}) as a linear combination of various terms which subsequently serve as the dynamical variables of the system. Above, we had the two variables $X$ and $Y$ and the simple constraint $X+Y=1$. As it was not possible to eliminate the Hubble function from the dynamical equations, it became necessary to introduce a third variable $z$ related to the Hubble function. This led to the two-dimensional system which was studied, with the second dimension representing the freedom in the parameters of $f$.

As we noted, the variables $Y$ and $z$ are not completely independent and it is possible to reduce the dimensionality of the system further, see also~\cite{Hohmann:2017jao}. To see this, let us first express $Y$ entirely in terms of $Q$ which gives
\begin{align}
    Y = \frac{f}{12 F H^2} = \frac{f}{2 F Q} := Y(Q) \,.
\end{align}
On the right-hand side we make it explicit that $Y$ can be seen as a function of $Q$. Using the chain rule we can write
\begin{align}
    \dot{Y} = \frac{\partial Y}{\partial Q} \dot{Q} \,,
\end{align}
while equation~(\ref{eqn:zq}) yields
\begin{align}
    \dot{z} = \frac{Q_0}{(Q_0+Q)^2} \dot{Q} \,.
\end{align}
Combining these previous two equations and the constraint $-\dot{X}=\dot{Y}$ we arrive at
\begin{align}
    -\dot{X} = \dot{Y} = \frac{\partial Y}{\partial Q} \frac{(Q_0+Q)^2}{Q_0} \dot{z} \,.
    \label{Xdotn1}
\end{align}
It is important to note that $Q$ can be re-written in terms of $z$. Consequently, one can now replace $\dot{X}$ on the left-hand side of~(\ref{eqn:gen1a}) using equation~(\ref{Xdotn1}). Next replace $X$ by $Y$, which is again a function of $z$, so that one arrives at a single equation in $z$. This slightly convoluted derivation gives the final equation
\begin{align}
    \frac{dz}{dN} = \frac{(1-z)(6z-(1-z)h(z))}{1+m(z)}\,,
    \label{eqn:ode1}
\end{align}
where we introduced the new function
\begin{align}
    h = \frac{f}{2F H_0^2} = \frac{3f}{F Q_0} \,,
\end{align}
which mirrors the form of the previous variable $Y$. Clearly, the entire dynamics for any given $f(Q)$ is now determined by the single first order ODE~(\ref{eqn:ode1}). For General Relativity with a cosmological constant $f(Q) = Q + 2\lambda_0$ one finds $m=0$, as stated earlier, and $h = 3z/(1-z)+\lambda_0/H_0^2$. The resulting first-order ODE becomes
\begin{align}
    \frac{dz}{dN} = (1-z)(3z-(1-z) \lambda_0/H_0^2)\,.
    \label{eqn:ode1GR}
\end{align}
It is straightforward to show that the dynamics encoded in~(\ref{eqn:ode1GR}) matches the previous discussion in two dimensions. Similar to before, the relation $H_\star=\sqrt{\lambda_0/3}$ emerges immediately from the stationary point of this equation. Here $H_\star$ stands for the value of $H$ where the right-hand side of~(\ref{eqn:ode1GR}) vanishes. This value may differ from $H_0$ which will determine the corresponding value of $z_\star$. It is, however, clear from~(\ref{eqn:ode1GR}) that the ratio $\lambda_0/H_0^2$ is the only free parameter in this ODE. 

Let us explain in different words why this dimensional reduction is possible. In the preceding sections there were three dynamical variables, which roughly related to $\rho$, $H$ and $f$. Together with the Friedmann constraint, one studies a system in two variables, $\{H,\rho\}$ say. However, since $f$ is a function of $H$ or $z$, the system is in fact `over-determined'. 

This is clearer when considering GR without a cosmological constant, $f(Q)=Q$. From the Friedmann constraint one immediately has $X=1/2$, and we're really dealing with a one-dimensional system (line), not a two-dimensional system. All other points on the two-dimensional phase space would be unphysical. Fig.~\ref{fig:GR1} can be seen as a collection of phase lines (i.e.~for each individual trajectory) for different values of the fixed parameter $\lambda_0$. In the remainder of this section and the succeeding ones, we remove $Y$ as dynamical variable and study the systems for fixed values of the constants in $f$. 

\subsection{$f(Q)$ gravity example -- Beltr{\'a}n et al.~model}

Let us apply this approach to the following model
\begin{align}
    f(Q)=Q-\lambda_p \frac{H_0^4}{Q} - 2\lambda H_0^2 \,,
\end{align}
which was suggested in~\cite{BeltranJimenez:2019tme} with partial results regarding a dynamical systems analysis. We also included a cosmological constant term $\lambda$ and introduced factors of $H_0$ such that both parameters $\lambda_p$ and $\lambda$ become dimensionless, which simplifies the subsequent discussion. Using the above formalism we developed, it is straightforward to compute explicit expressions for $m(z)$ and $h(z)$. When these are put into the first order ODE~(\ref{eqn:ode1}) one finds
\begin{align}
    \frac{dz}{dN} = \frac{3z(1-z)z\bigl[4z(\lambda+z(3-\lambda))+\lambda_p(1-z)^2\bigr]}
    {z^2(12-\lambda_p)+2z\lambda_p-\lambda_p} =: b(z)\,.
    \label{eqn:ode1Bel}
\end{align}
For $\lambda_p < 0$ the denominator cannot vanish and we are dealing with a regular ODE. When setting $\lambda_p = 0$ we are back at GR with a cosmological constant as discussed above. When $\lambda_p > 0$ we find that the denominator of~(\ref{eqn:ode1Bel}) can become zero when
\begin{align}
    z_s = \frac{\lambda_p - 2\sqrt{3\lambda_p}}{\lambda_p -12} \,.
\end{align}
We note that $0 < z_s \leq 1$ for all $\lambda_p>0$, which means that for positive $\lambda_p$ one always finds a singular denominator. Other than $z=0$ and $z=1$, the numerator can also vanish, giving rise to up to two critical points. These are 
\begin{align}
    z_c = \frac{\lambda_p-2\lambda\pm2\sqrt{\lambda^2-3\lambda_p}}
    {\lambda_p-4\lambda+12} \,,
\end{align}
which can give two distinct critical points in the permissible range if $\lambda_p>0$ and $\lambda < -\sqrt{3\lambda_p}$. A direct calculation shows that $b'(z_c)=3$, which means that any critical point $z_c$ will always be unstable for any choice of parameter.

\begin{figure}[!htb]
    \centering
    \includegraphics[width=0.5\textwidth]{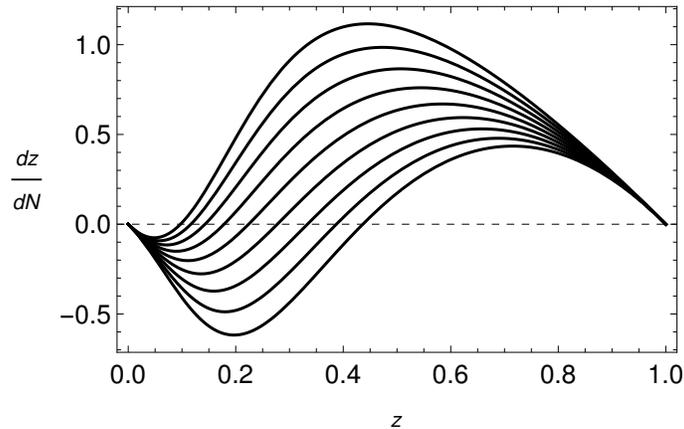}
    \caption{Phase space of the model given by $f(Q)=Q-\lambda_p \frac{H_0^4}{Q} - 2\lambda H_0^2$. $\lambda_p < 0$ is assumed. The value of $\lambda$ increases from the bottom to the top with values $\lambda=-2,-1.5,\ldots,2$.}
    \label{fig:Bel1}
 \end{figure}

The typical phase space for models with $\lambda_p < 0$ is shown in Fig.~\ref{fig:Bel1} for varying value of the cosmological constant $\lambda$. All solutions evolve either towards $z=0$ or $z=1$, depending on the initial conditions.

 The following Fig.~\ref{fig:Bel2} shows the typical phase space plot for $\lambda_p > 0$. The dashed (red) vertical line represents the singular line of the ODE. Since this line corresponds to a fixed value of $z=z_s$, we have that the Hubble parameter also approaches a constant values $H(z) \to H(z_s)=H_s$. However, we note that $\lim_{z\to z_s} H'(z) \to \pm \infty$, depending on the initial conditions.

\begin{figure}[!htb]
    \centering
    \includegraphics[width=0.5\textwidth]{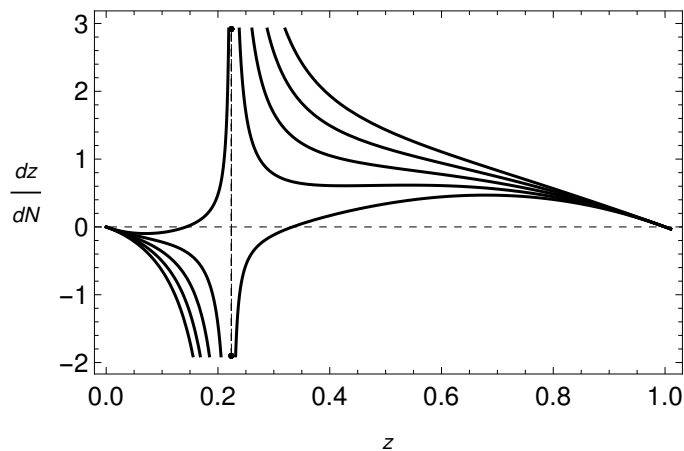}
    \caption{Phase space of the model given by $f(Q)=Q-\lambda_p \frac{H_0^4}{Q} - 2\lambda H_0^2$. $\lambda_p > 0$ is assumed. The value of $\lambda$ increase from the bottom to the top with values $\lambda=-2,-1,\ldots,2$.}
    \label{fig:Bel2}
 \end{figure}

In the following we will consider models with matter and radiation. These models will be two-dimensional and will be conveniently studied using dynamical systems techniques. This will be analogous to the approach taken in GR, where models with matter and radiation can also be studied using a two-dimensional phase space.

\section{Dynamical systems -- alternative formulation}

\subsection{General setup with matter and radiation}

Generalising the formulation in Section~\ref{sec:dimred}, where the dimensionality of the system is one fewer than in the previous discussions, we now look at the case with two fluids, matter $\rho_m$ ($w= 0$) and radiation $\rho_r$ ($w= 1/3$). The cosmological equations can be written as
\begin{align} \label{cosmo1}
  -\frac{f}{6H^2} - \frac{\rho_m}{3H^2} - \frac{\rho_r}{3H^2}+ 2 F &= 0 \,, \\
  \label{cosmo2}
  \rho_m + \frac{4 \rho_r}{3} + 2 \dot{H} (F + 12 H^2 F') &= 0 \,.
\end{align}
Let us introduce the following dynamical variables, which we note differ from (\ref{XY1}) by having no $F$ in the denominator of the fluid variables
 \begin{align}
    X_1 = \frac{\rho_m}{3 H^2 } \,, \qquad X_2 = \frac{\rho_r}{3 H^2} \,, \qquad Z = \frac{6H^2}{6H_0^2 +6H^2} \,.
    \label{X1X2Z}
\end{align}
The first two variables are the canonical \emph{EN-variables} commonly used in cosmology, and $Z$ is the same as $z$ before. Note that none of the variables~(\ref{X1X2Z}) depend on the choice of function $f$. This is possible because any function of $H$ can be written in terms the variable $Z$, which will only enter the dynamics via the Friedmann constraint itself. The benefit is that $X_1$ and $X_2$ are now non-negative and take a familiar form.

The Friedmann constraint can then be rewritten solely in terms of the new variables
\begin{align} \label{2fluidF}
    0 &= X_1 + X_2 + \frac{f}{6H^2} - 2 F \nonumber \\
    X_1 + X_2 &=  \big(1-\frac{1}{Z}\big) \frac{f}{6 H_0^2} + 2 F \ ,
\end{align}
where in the final line we have treated $f = f({6 H_0^2 Z}/{(1-Z)})$ as a function of $Z$, and similarly for $F$. These functions can always be seen in two different ways: either as functions of the scalar that determines the modified theory of gravity; or as functions of the Hubble parameter, which we express in terms of $Z$. Writing $f$ and $F$ in this way means that the expression for $X_1 + X_2$ is completely in terms of $Z$ once $f$ is specified. We can therefore use this equation to eliminate either $X_1$ or $X_2$, making the phase space two-dimensional. Let us choose to eliminate $X_1$ and consider the evolution of $\{X_2,Z\}$,
\begin{align} \label{X2 eq long}
    \frac{d X_2}{d N} &= - \frac{X_2 \left((Z-1)^2 f + 2 H_0^2 Z \left( X_2 (Z-1) + 2(Z-1) F + 48 H_0^2 Z F' \right) \right) }{2 H_0^2 Z \big((1-Z)F + 12 H_0^2 Z F' \big)} \, , \\ \label{Z eq long}
    \frac{d Z}{d N} &= -\frac{(Z-1)^2 \left((Z-1)f + 2 H_0^2 Z (X_2 + 6 F ) \right)}{2 H_0^2 \big((1-Z)F + 12 H_0^2 Z F' \big)} \, ,
\end{align}
where we have made use of (\ref{2fluidF}). Note that $f$ has dimensions $H_0^2$, $F$ is always dimensionless and $F'$ has units of $H_0^{-2}$. Consequently, equations~(\ref{X2 eq long})--(\ref{Z eq long}) are dimensionless, with the numerators and denominators all having equal powers of $H_0$.

Looking at equations (\ref{X2 eq long})--(\ref{Z eq long}), one can make some general statements about the systems. For convenience, we introduce the functions
\begin{align} \label{m(Z)}
    m(Z) &:= \frac{12 H_0^2 Z F'}{(1-Z) F} \,, \\ \label{n(Z)}
    n(Z) &:= \frac{f (1-Z)}{6H_0^2 Z F} \,,
\end{align}
such that the dynamical equations can be compactly rewritten as
\begin{align} 
    \label{X2 mn}
    \frac{d X_2}{d N} &= \frac{X_2 \left(2-3n-4m+ X_2/F \right)}{1+m} \,, \\ 
    \label{Z mn}
    \frac{d Z}{d N} &= \frac{Z(Z-1) \left(X_2 - 3(n-2)F \right)}{(1+m)F} \,.
\end{align}
Again we remind the reader that here $f$ is defined as $f = f(\frac{6 H_0^2 Z}{1-Z})$. Note that once $f$ is specified, we are dealing with a closed system of equations which can be studied. However, it turns out that some properties of this system are independent of this function and hence are valid for all models, making our approach very broad.

\subsection{Fixed Points}

The fixed points of the system can be divided into two families of solutions, one with $X_2=0$ and the other with $X_2=(4m+3n-2) F =: X_2^*$, provided they exist and are within the physical phase space (see discussion below). For the case where $X_2=0$, the points with $Z=Z^*$ are solutions of the algebraic equation 
\begin{align} 
    \label{eq Z*}
    \frac{3Z(Z-1)(n(Z)-2)}{1+m(Z)}=0 \,.
\end{align}
For the $X_2=X_2^*$ case, the $Z$ coordinates are $Z=0$ or $Z=1$. The $X_2$ coordinates must then be evaluated at $Z=0$ or $Z=1$ because $X_2^*$ is function of $Z$.

These results have been collected in Table~\ref{tab2} with the P's representing points along the $X_2=0$ line with $Z=Z^*$ satisfying~(\ref{eq Z*}). The special cases of $Z^*$ when $n(Z)=2$ and $m(Z)\rightarrow \infty$ are labelled as P$_n$ and P$_m$ respectively. Note that $m(Z)$ can diverge when $F \rightarrow 0$, and $F$ appears in both~(\ref{X2 mn}) and~(\ref{Z mn}) and the denominator of $n(Z)$~(\ref{n(Z)}). However, $F \rightarrow 0$ alone does not represent a critical point of the system. For this reason, we also require $n(Z)$ to be finite whilst $m(Z)\rightarrow \infty$. In other words, we require that $F' \rightarrow \infty$ or that  both $f \rightarrow 0$ and $F \rightarrow 0$ for some value(s) of $Z$. 

The points A and B are solutions with $X_2=X_2^*$ evaluated at $Z=0$ and $Z=1$. Lastly, just like in the previous analysis, there exists a singular line L1 if there exists values $Z=Z_c$ such that $m(Z_c)=-1$. There also exists a singular line L2 when $n(Z) \rightarrow \infty$ with $m(Z)$ finite. This means that either $f \rightarrow \infty$ or that both $F \rightarrow 0$ and $F' \rightarrow 0$ such that $n(Z)$ diverges but $m(Z)$ does not, similar to the conditions for P$_m$.

The corresponding values of $X_1$ determined by the Friedmann constraint have also been included. The final column states the specific conditions for each point, and it is assumed that all variables $X_1$, $X_2$, $Z$ must be finite for existence. Note that the fixed points are not necessarily unique and in many cases they coincide with one another, see Table~\ref{tab2}.

\begin{table}[htb!]
    \centering
    \begin{tabular}{c|c|c|c|p{5cm}}
         Point/Line & $X_2$ & $Z$ & $X_1$ & conditions \\ \hline\hline
         P$_1$ & $X_2=0$ & $Z=0$ & $X_1=(2-n(0))F(0)$ & $m(0) \neq -1$, $n(0) \neq 2$, both finite  \\
           P$_2$ & $X_2=0$ & $Z=1$ & $X_1=(2-n(1)) F(+\infty)$ & $m(1) \neq -1$, $n(1) \neq 2$, both finite  \\
          P$_n$ & $X_2=0$ & $Z=Z^*$ & $X_1 = 0$ & $n(Z^*)=2$, $m(Z^*)$ finite \\
           P$_m$ & $X_2=0$ & $Z=Z^{*}$ & $X_1=(2-n(Z^{*}))F(Z^{*})$ & $m(Z^{*}) \rightarrow \infty$, $n(Z^*)$ finite \\
          A & $X_2=X_2^*|_{Z=0}$ & $Z=0$ & $X_1=-4(m(0)+n(0)-1)F(0)$ & $X_1$, $X_2$ well-defined \\ 
          B & $X_2=X_2^*|_{Z=1}$ & $Z=1$ & $X_1=-4(m(1)+n(1)-1)F(+\infty)$ & $X_1$, $X_2$ well-defined \\
           L1 & any & $Z=Z_c$ & $X_1= -X_2 +(2-n(Z_c))F(Z_c)$ & $m(Z_c) = -1$, $n(Z_c)$ finite \\
           L2 & any & $Z=Z_d$ & $X_1=-X_2 + (2-n(Z_d))F(Z_d)$ & $n(Z_d) \rightarrow \infty$, $m(Z_d)$ finite \\
    \end{tabular}
    \caption{Table of fixed points of system~(\ref{X2 mn})--(\ref{Z mn}) for arbitrary function $f$.}
    \label{tab2}
\end{table}

\subsection{Physical Parameters}

Using the formulation above (\ref{m(Z)})--(\ref{Z mn}) along with the cosmological field equations, the deceleration parameter can be expressed in terms of the new variables 
\begin{align}
    q := -\frac{\ddot{a}a}{\dot{a}^2} = -1-\frac{\dot{H}}{H^2} = -1 - \frac{(\frac{3}{2}X_1+2X_2)(n-2)}{(X_1+X_2)(1+m)} \,.
\end{align}
The value of $q$ for each of the fixed points is given in Table~\ref{tab3}. For the points P$_n$, P$_m$, A and B one obtains fixed values of $q$, which are model independent. For the lines L1 and L2, the deceleration parameter diverges. The points P$_1$ and P$_2$ depend on the functions $m$ and $n$ evaluated at their respective $Z$ values.

The effective equation of state parameter $w_{\rm{eff}}$ can be expressed in a similar form
\begin{align}
  w_{\rm{eff}} := \frac{p_{\rm{tot}}}{\rho_{\rm{tot}}} = 
  -1 + \frac{(X_1+\frac{4}{3}X_2)(n-2)}{(X_1+X_2)(1+m)} \,,
\end{align}
where the total energy density is $\rho_{\rm{tot}}= \rho_{m}+\rho_{r} + \rho_{f}$ and $\rho_{f}$ represents the extra terms in (\ref{cosmo1}) that do not appear in the standard Friedmann equation in General Relativity. Similarly, the total pressure is $p_{\rm{tot}}= p_{m}+p_{r} + p_{f}$ with $p_{f}$ representing the additional terms in the acceleration equation~(\ref{cosmo2}). The values of $w_{\rm{eff}}$ evaluated at the critical points are also given in Table~\ref{tab3}. Lastly, the values of the Hubble parameter $H(t)$ are also included, provided they are well-defined. 

\begin{table}[htb!]
    \centering
    \begin{tabular}{c|c|c|c}
         Point/Line & $q$ & $w_{\rm{eff}}$ & $H(t)$ \\ \hline\hline
         && \\[-8pt]
       P$_1$ & $\displaystyle\frac{4-2m(0)-3n(0)}{2+2m(0)}$ & $\displaystyle-\frac{-1+m(0)+n(0)}{1+m(0)}$ & $H(t) = 0$ \\[8pt]
         P$_2$ & $\displaystyle\frac{4-2m(1)-3n(1)}{2+2m(1)}$ & $\displaystyle-\frac{-1+m(1)+n(1)}{1+m(1)}$ & $H(t) \rightarrow \pm \infty$ \\
         P$_n$ & $-1$ & $-1$ & $H(t) =$ const. \\
         P$_m$ & $-1$ & $-1$ & $H(t) =$ const. \\
         A & $1$ & $1/3$ & $H(t) = 0$ \\
         B & $1$ & $1/3$ & $H(t) \rightarrow \pm \infty$ \\
         L1 & $\pm \infty$ & $\pm \infty$ & not defined \\
         L2 & $\pm \infty$ & $\pm \infty$ & not defined \\
    \end{tabular}
    \caption{Physical quantities at critical locations}
    \label{tab3}
\end{table}

\begin{figure}[!htb]
    \centering
    \includegraphics[width=0.5\textwidth]{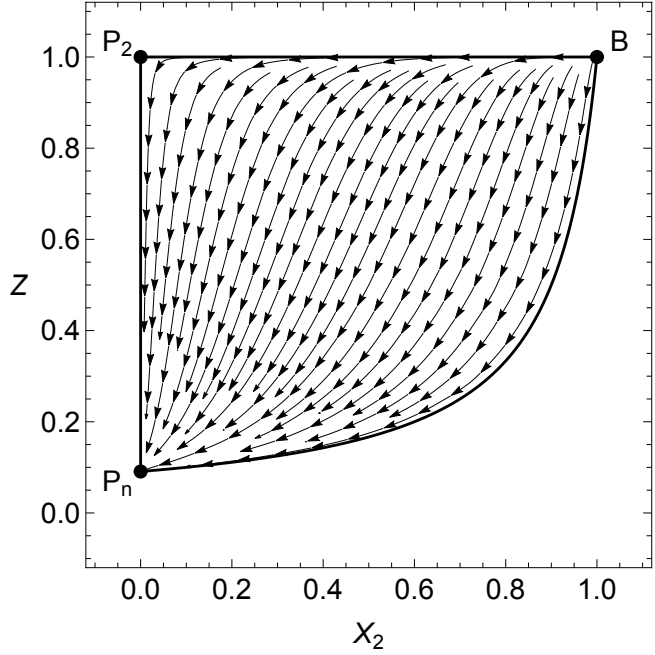}
    \caption{Phase space for General Relativity with $f(Q)=Q + 6 \Lambda H_0^2$ with $\Lambda >0$.}
    \label{fig:gr2fluid}
\end{figure}

Note again that we require $X_1$ and $X_2$ to be non-negative and that $0 \leq Z \leq 1$. In order to enforce these conditions on the physical phase space, we must use equation~(\ref{2fluidF}) for a given $f$ to constrain $X_2$ and $Z$. Then one can simply read off the fixed points from Table~\ref{tab2} (that satisfy their appropriate existence conditions) and determine whether they are located within the physical phase space. For example, for $f(Q)=Q$ equation~(\ref{2fluidF}) simplifies to $X_1+X_2 =1$ and the physical phase space in $\{X_2,Z\}$ is the unit square. On the other hand, for General Relativity with a cosmological constant $f(Q)=Q + 6 \Lambda H_0^2$ and $\Lambda > 0$ one instead has $X_1 + X_2 = 1 + \Lambda - \Lambda/Z$. The physical phase space is then the region satisfying $Z \geq \Lambda /(1-X_2+ \Lambda)$ and $0 \leq Z \leq 1$, $X_2 \leq 1$, see Fig.~\ref{fig:gr2fluid}. Only the fixed points in Table~\ref{tab2} that lie within the region constrained by~(\ref{2fluidF}) are physical. In Fig.~\ref{fig:gr2fluid} we see the points B, P$_2$ and P$_n$, representing the radiation repeller, matter saddle and de Sitter attractor, respectively.

Before studying other concrete models in this formulation, we quickly look at the linear stability of the system in full generality. 

\subsection{Stability Analysis}

As in the previous section, the Jacobian matrix or stability matrix can be analysed at the fixed points in Table~\ref{tab2} to make some general statements about the stability of critical points. The necessary existence conditions are also assumed, but care must be taken with the derivative of the function $m(Z)$, which will contain third derivatives of $f$ (the derivatives of $n(Z)$ can be rewritten in terms of $m$ and $n$). It is often necessary to assume that $F''$, which appears in the numerator of the elements of the Jacobian matrix, remains finite at the fixed point in order to determine the stability for a general $f$.

For P$_1=(0,0)$ one obtains $\lambda_1 = (2-4m(0)-3n(0)) / (1+m(0))$ and $\lambda_2 = (3n(0)-6) / (1+m(0))$, which has at least one negative eigenvalue. Conditions on $m(Z)$ and $n(Z)$ could then be imposed to determine its stability. The second point P$_2=(0,1)$ has the same first eigenvalue $\lambda_1=(2-4m(1)-3n(1)) / (1+m(1))$ and its second eigenvalue is $\lambda_2 = -(3n(1)-6) / (1+m(1))$, where the close similarities should be noted but with $m$ and $n$ evaluated at $Z=1$. Unsurprisingly, with $m$ and $n$ often undefined at these points, there is little to be said about the stability of these points in general, see again their definitions~(\ref{m(Z)}) and~(\ref{n(Z)}).

For the point P$_n=(0,Z^*)$ with $n(Z^*)=2$ one gets constant eigenvalues of $\lambda_1=-4$ and $\lambda_2=-3$, meaning this point is always stable when it exists. This is consistent with its interpretation as being a late-time de Sitter attractor with $q=-1$. The stability of the point P$_m=(0,Z^*)$ with $m(Z^*) \rightarrow \infty$ depends on the third derivatives of $f$, so one cannot make definitive claims about its stability. However, if one checks that both $n(Z^*)$ and $F''(Z^*)$ stay finite whilst $m(Z^*) \rightarrow \infty$, the eigenvalues will also be $\lambda_1=-4$ and $\lambda_2=-3$, indicating stability.

Point A $=(X_2^*,0)$, with $X_2^*$ defined previously, has eigenvalues $\lambda_1=-4$, $\lambda_2=-(2-4m(0)-3n(0))/(1+m(0))$. The second eigenvalue is just minus the first of point P$_1$'s, therefore both points cannot be stable. For point B the eigenvalues depend on the higher derivatives of $f$, so again, the stability cannot be determined without specifying a model $f$. If all derivatives of $f$ stay finite at $Z=1$ it is possible to at least conclude that this point must be unstable, as the eigenvalues cannot both be negative.

For the lines L1 and L2, all components of the Jacobian diverge, making an analysis is not possible without defining a specific $f(Q)$. Note that in this analysis we have again assumed that all quantities are finite unless otherwise stated, such that none of the components of the stability matrix diverge at the fixed points considered above.

\subsection{Revisiting the Beltr\'an et al.~model with two fluids} 
\label{subsection: beltran}

The model we are considering now is given by
\begin{align} 
    \label{beltran lagrangian}
    f(Q)=Q-\lambda_p \frac{H_0^4}{Q} - 2\lambda H_0^2 \,,
\end{align}
and we assume the presence of two fluids, matter and radiation. In this case, the dynamical equations become
\begin{align} \label{beltran eqs1}
  \frac{d X_2}{dN} &= -\frac{X_2 \left(2Z(6\lambda - 7 \lambda_p) + 7 \lambda_p + Z^2 \left(12 X_2 - 12 \lambda + 7 \lambda_p - 12\right) \right)}
       {Z^2(\lambda_p - 12) +\lambda_p - 2 Z \lambda_p} \,, \\
     \label{beltran eqs2}
    \frac{d Z}{dN} &= -\frac{3Z(Z-1) \big(Z(4\lambda-2 \lambda_p) + \lambda_p + Z^2(12+4 X_2 - 4 \lambda + \lambda_p) \big)}{Z^2(\lambda_p -12) + \lambda_p - 2 Z \lambda_p} \,,
\end{align}
and the functions $m(Z)$ and $n(Z)$ are given by
\begin{align} 
    \label{beltran m}
    m(Z) &= -\frac{4 \lambda_p (Z-1)^2}{\lambda_p - 2 Z \lambda_p + Z^2 (36 + \lambda_p)}\,, \\ 
    n(Z) &= \frac{Z^2(36+12 \lambda - \lambda_p) - \lambda_p + 2Z (\lambda_p- 6 \lambda)}{\lambda_p - 2 Z \lambda_p + Z^2 (36+\lambda_p)} \label{beltran n} \,.
\end{align}
We will assume that $\lambda_p \neq 0$ as this would reduce the model back to GR with a cosmological constant term. The sign of the parameter $\lambda_p$ determines whether the phase space is naturally compact or not, so we will look at these two cases separately.

From Table~\ref{tab2} the existence of generic fixed points can be examined, but the Friedmann constraint (\ref{2fluidF}) will need to be used to determine which are physically relevant. Both conditions for P$_1$ and P$_2$ are satisfied. For the point P$_n$ with $n(Z)=2$, the condition is that either $\lambda_p <0$ or $\lambda_p \geq 0$ with $\lambda^2 - 3 \lambda_p \geq 0$. In fact, one can also see that for $\lambda_p <0$ the algebraic equation $n(Z)=2$ has one solution for $0 \leq Z \leq 1$, except in the special case when $\lambda$ is chosen such that $n(Z)$ does not diverge for any $Z$ in the physical range (see Appendix~\ref{Appendix Beltran} for details, where in this pathological case the point P$_n$ is replaced by P$_m$). For $\lambda_p > 0$ the equation $n(Z)=2$ can either have zero, one, or two solutions within $0 \leq Z \leq 1$ depending on whether $\lambda$ is greater than, equal to, or less than $-\sqrt{3\lambda_p}$.

The point P$_m$ does not exist -- except in the special case mentioned above which is discussed in Appendix~\ref{Appendix Beltran} -- because $m(Z)$ and $n(Z)$ have the same denominators, meaning they both diverge for the same values of $Z$. Evaluating $X_2^{*}$ at $Z=0$ and $Z=1$ reveals that point A has $X_2 \rightarrow \pm \infty$, whilst point B has $X_2 =1$. Lastly, the line L1 with $m(Z)=-1$ exists for $\lambda_p > 0$ with at most one solution in the physical $Z$ range, whereas L2 does not exist because $n(Z)$ cannot diverge with $m(Z)$ staying finite. Next we will look at the phase space plots for this model.

\begin{figure}[!htb]
     \centering
     \begin{subfigure}[b]{0.4\textwidth}
         \centering
         \includegraphics[width=\textwidth]{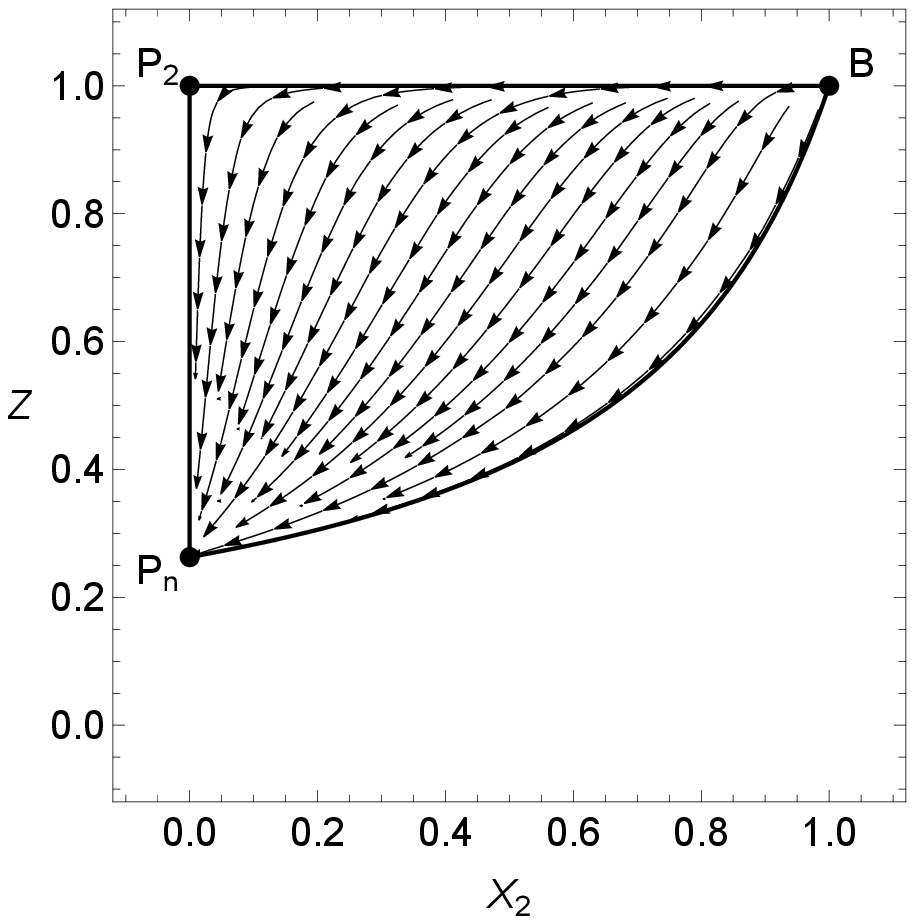}
         \caption{$\lambda_p < 0$ and $\lambda < 0$.}
         \label{fig:beltran1}
         \vspace{5mm}
     \end{subfigure} \hspace{3mm}
    \begin{subfigure}[b]{0.4\textwidth}
         \centering
     \includegraphics[width=\textwidth]{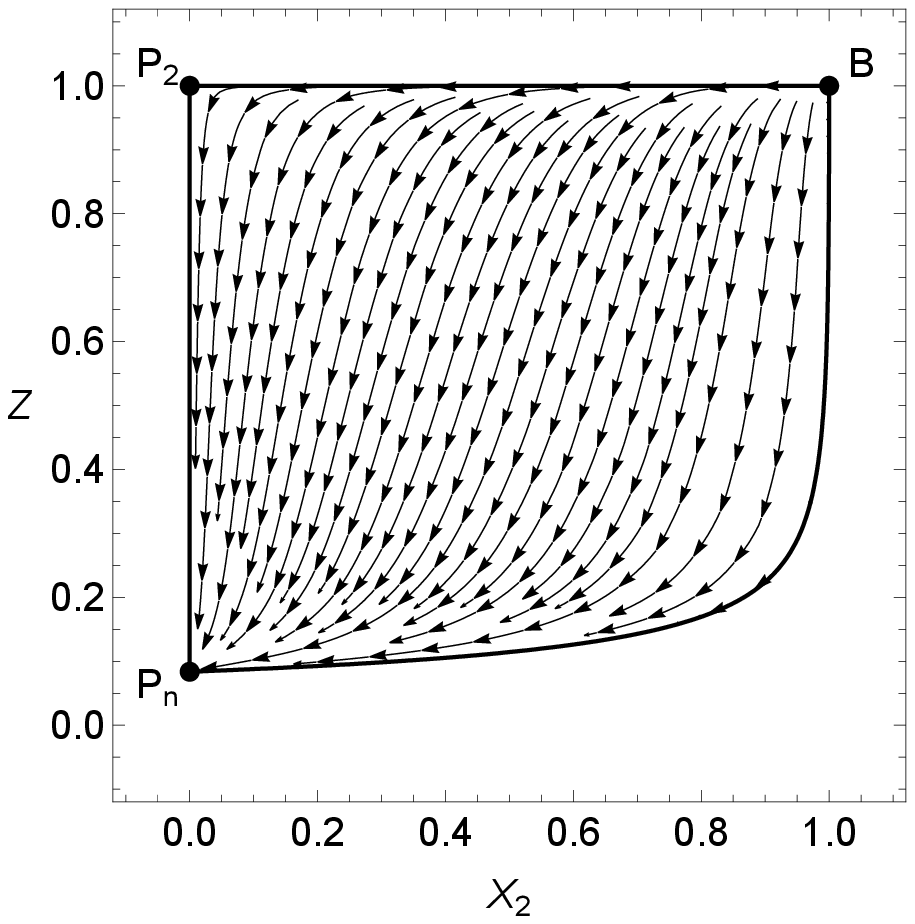}
         \caption{$\lambda_p < 0$ and $\lambda = 0$.}
         \label{fig:beltran1b}
         \vspace{5mm}
     \end{subfigure}
     \hfill
     \begin{subfigure}[b]{0.75\textwidth}
         \centering
         \includegraphics[width=\textwidth]{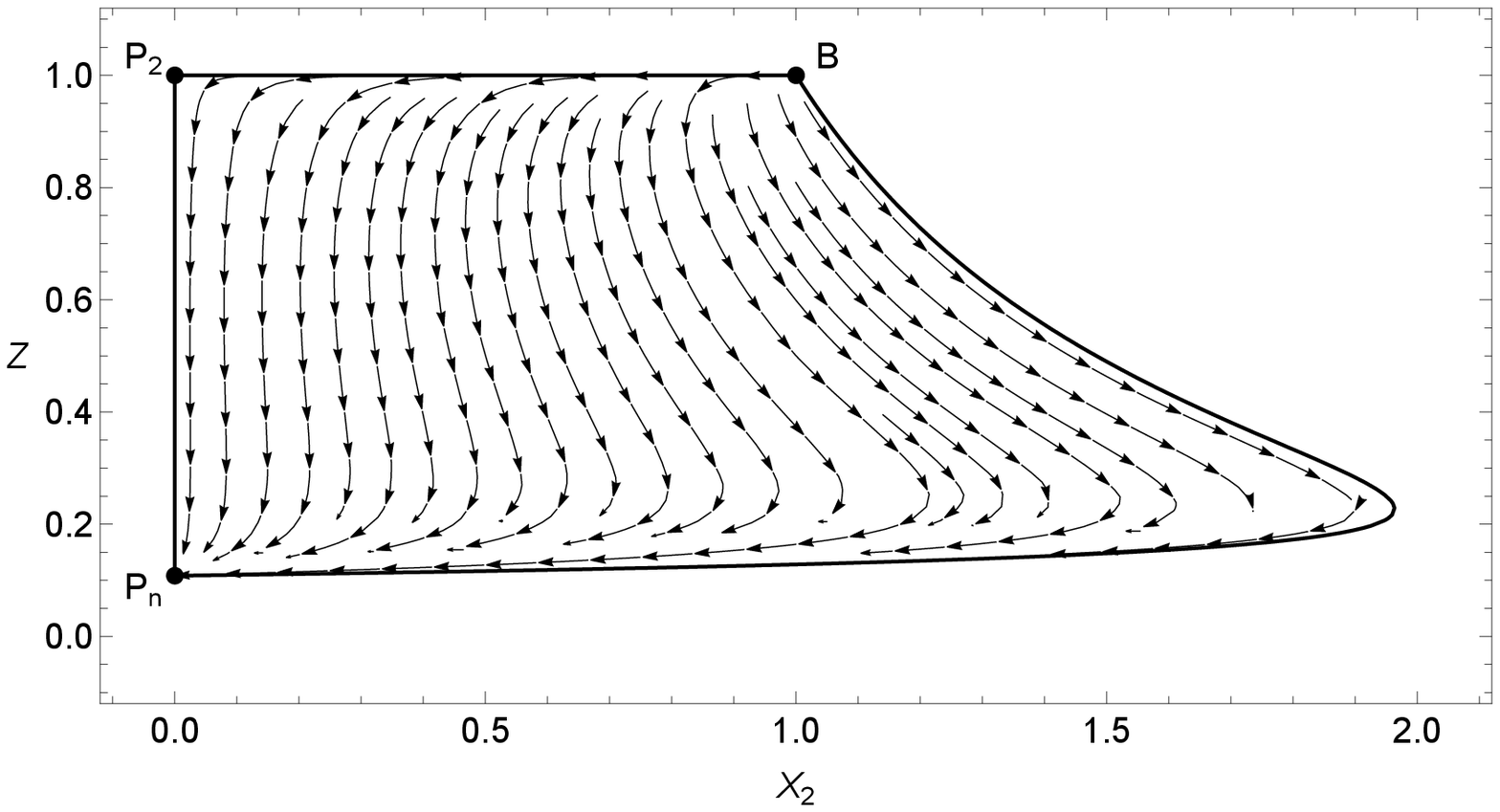}
         \caption{$\lambda_p < 0$ and $\lambda > 0$.}
         \label{fig:beltran2}
     \end{subfigure}
        \caption{Phase space for the Beltran model (\ref{beltran lagrangian}) with $\lambda_p <0$.}
        \label{fig:beltran sub1}
\end{figure}

First, assume $\lambda_p <0 $ and $\lambda \leq 0$. In this case, the model is compact and $X_1$ and $X_2$ are bounded between $[0,1]$, which can be deduced from~(\ref{2fluidF}). There is also the additional constraint from~(\ref{2fluidF}) for the physical region to satisfy the inequality
\begin{align} 
    \label{beltran friedmann}
    X_1 = \frac{Z(4\lambda-2\lambda_p) + Z^2(12-12 X_2 -4 \lambda + \lambda_p) + \lambda_p}{12 Z^2} \geq 0 \, ,
\end{align}
which must hold for any values of the parameters. The phase portraits for $\lambda <0$ and $\lambda =0$ are given in Fig.~\ref{fig:beltran1} and Fig.~\ref{fig:beltran1b}. The fixed points B, P$_2$ and P$_n$ represent the radiation dominated repeller $\Omega_{r}=1$, matter dominated saddle point $\Omega_{m}=1$ and de Sitter attractor $\Omega_{\Lambda}=1$, respectively. This can be deduced from the the definitions of $X_2$ and $Z$ at the fixed points, and by using the Friedmann constraint to determine the value of $X_1$ ($\Omega_{m}$). From Table~\ref{tab3} the effective equation of state parameter can be read off at the points B and P$_n$ as $w_{\rm{eff}}=1/3$ and $w_{\rm{eff}}=-1$. For the matter saddle point P$_2$ one easily finds $w_{\rm{eff}}=0$ as expected.
All orbits start at B and end at P$_n$, with some trajectories attracted towards the matter saddle P$_2$. The qualitative features are very similar to those of GR with a positive cosmological constant, shown in Fig.~\ref{fig:gr2fluid}. 

When $\lambda_p$ is negative but $\lambda > 0$ the phase space is still compact but no longer bounded between $[0,1]$ in the fluid variables $X_1$ and $X_2$. The late time de Sitter point P$_n$ at $(0,Z^*)$ remains the only global late-time attractor of the physical phase space, and the other two fixed points are the same. However, trajectories beginning at the past-attractor, point B, can instead follow trajectories in the the positive $X_2$ direction before terminating at the future-time attractor P$_n$, shown in Fig.~\ref{fig:beltran2}. This is because equation~(\ref{beltran friedmann}), which defines the physical phase space, changes when the parameters of the model change. 

For the case when $\lambda_p > 0$ the situation differs. Firstly, the variables $X_1$ and $X_2$ are no longer bounded from above, meaning the phase space is not compact. It is instead constrained by the inequality 
\begin{align}
    0 \leq X_2 \leq \frac{Z(4 \lambda - 2 \lambda_p) + \lambda_p + Z^2 (12 - 4 \lambda + \lambda_p)}{12 Z^2} \ ,
\end{align}
along with $0 \leq Z \leq 1$, as usual. The phase space is then infinite in the positive $X_2$ direction as $Z$ tends to zero. 

\begin{figure}[!htp]
     \centering
     \begin{subfigure}[b]{0.45\textwidth}
         \centering
         \includegraphics[width=\textwidth]{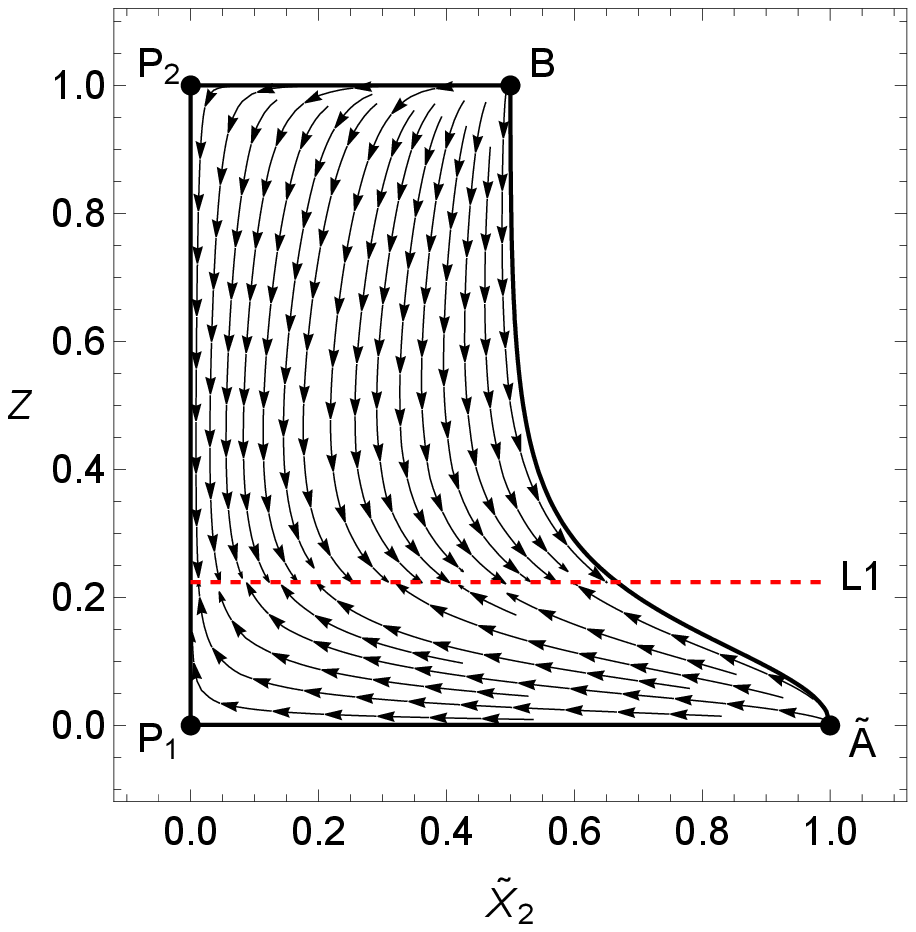}
         \caption{$\lambda_p > 0$ and $\lambda > - \sqrt{3 \lambda_p}$.}
         \label{fig:beltran3}
         \vspace{6mm}
     \end{subfigure}  \hspace{2mm}
     \begin{subfigure}[b]{0.45\textwidth}
         \centering
         \includegraphics[width=\textwidth]{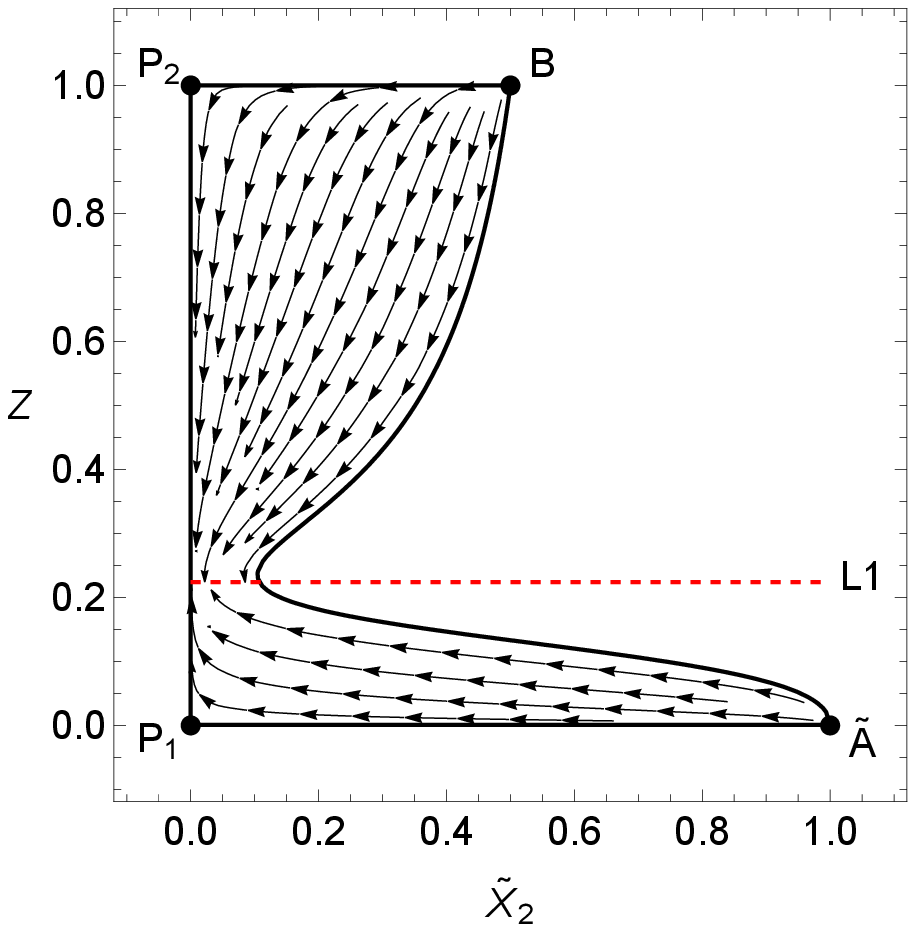}
         \caption{$\lambda_p > 0$ and $\lambda > - \sqrt{3 \lambda_p}$.}
         \label{fig:beltran4}
         \vspace{6mm}
     \end{subfigure}
     \begin{subfigure}[b]{0.45\textwidth}
         \centering
         \includegraphics[width=\textwidth]{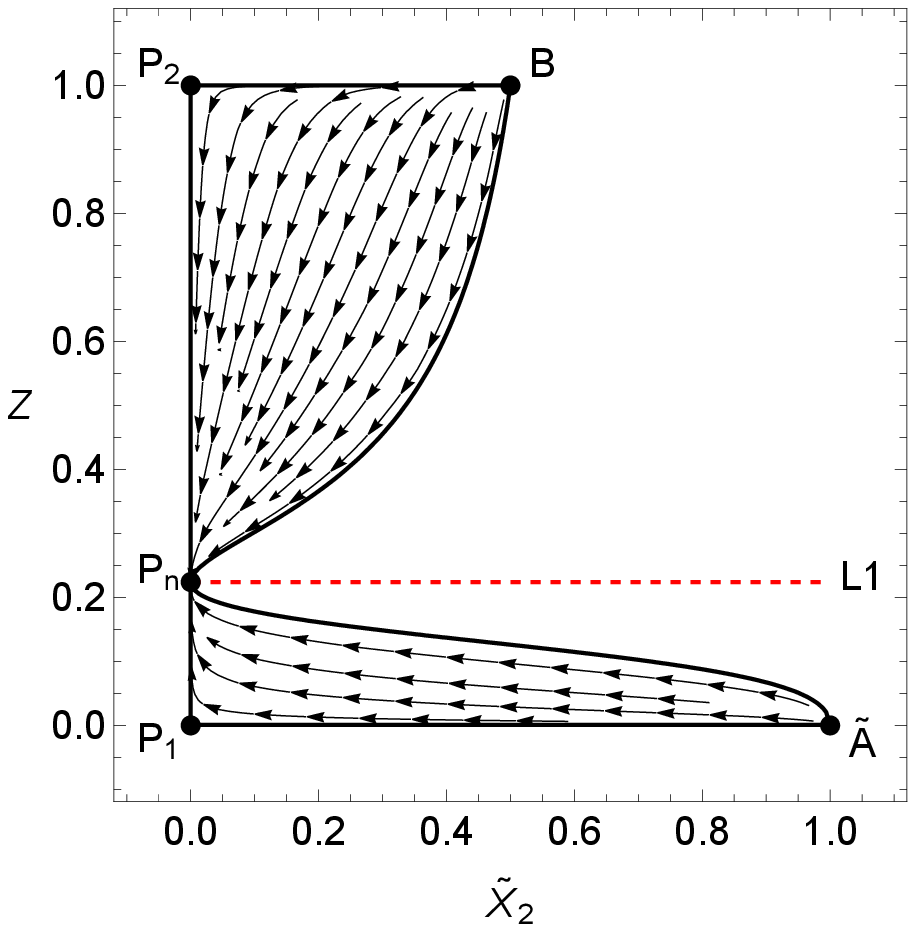}
         \caption{$\lambda_p > 0$ and $\lambda = - \sqrt{3 \lambda_p}$.}
         \label{fig:beltran5}
     \end{subfigure}  \hspace{2mm}
     \begin{subfigure}[b]{0.45\textwidth}
         \centering
         \includegraphics[width=\textwidth]{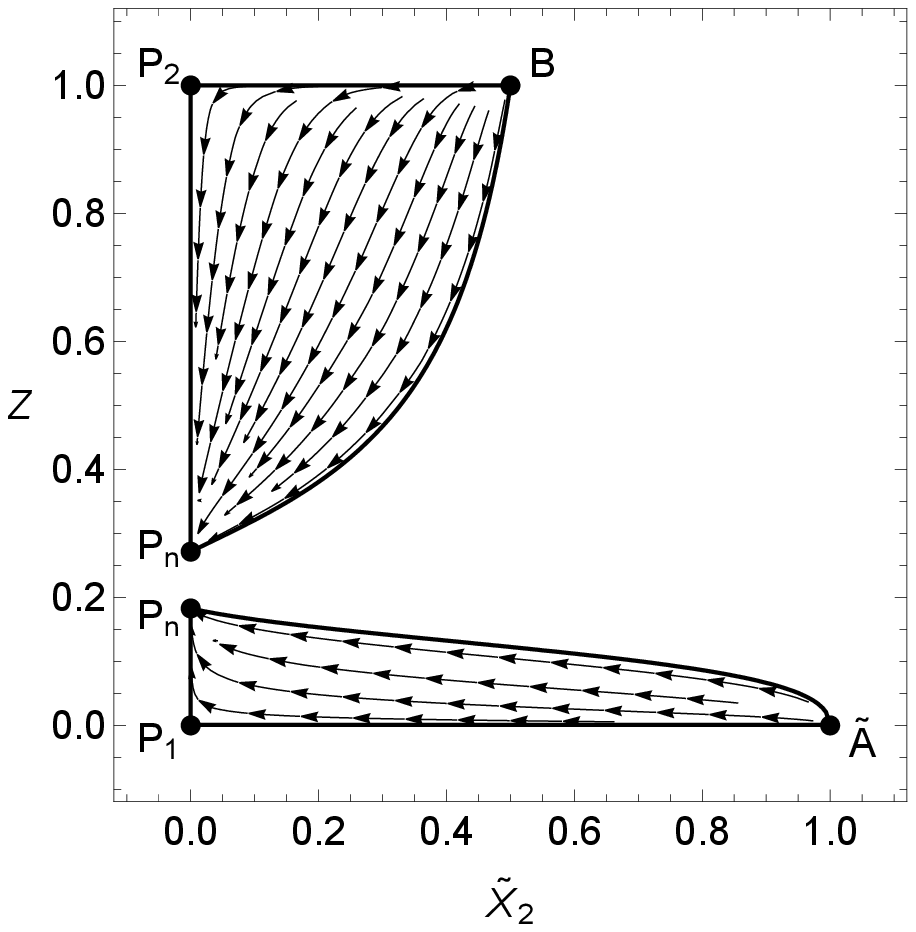}
         \caption{$\lambda_p > 0$ and $\lambda < - \sqrt{3 \lambda_p}$.}
         \label{fig:beltran6}
     \end{subfigure}
        \caption{Compactified phase space for the Beltran model (\ref{beltran lagrangian}) with $\lambda_p >0$.}
        \label{fig:beltran sub2}
\end{figure}

Analogous to the 1D case (Fig.~\ref{fig:Bel2}), the denominator of equations~(\ref{beltran eqs1}) and~(\ref{beltran eqs2}) vanishes when the $Z$ coordinate takes the fixed value $Z=Z_c (\lambda_p)$ such that $m(Z_c)=-1$. This corresponds to the line L1 with $Z$ coordinate $Z_c = (-2 \sqrt{3 \lambda_p} + \lambda_p) / (\lambda_p -12)$. If $\lambda > - \sqrt{3 \lambda_p}$ the physical region is connected, whereas for $\lambda \leq - \sqrt{3 \lambda_p}$ the space bifurcates into two disconnected parts at the point point $\{0,Z=Z_c\}$. However, for all $\lambda$, trajectories are always confined to regions either above or below the line L1 at $Z=Z_c$. This will be shown clearly in the compactified phase portraits, Fig.~\ref{fig:beltran sub2}.

In order to compactify the phase space we introduce the variable 
\begin{align}
    \tilde{X}_2 = \frac{X_2}{1+X_2} \ ,
\end{align}
such that $0 \leq \tilde{X}_2 \leq 1$. This is what we referred to earlier as our \emph{poor man's version} of compactification. The new dynamical system in $\{\tilde{X}_2,Z\}$ can then be found by taking the derivative of $\tilde{X}_2$ with respect to $N$ and using equations~(\ref{beltran eqs1})--(\ref{beltran eqs2}). Similarly, the Friedmann constraint is rewritten with $\tilde{X_2}$ leading to the compactified phase spaces given in Fig.~\ref{fig:beltran sub2}. Note that the fixed points in Table~\ref{tab2} and Table~\ref{tab3} are still valid in the original variables $X_2$ and $Z$, therefore they will also represent critical points in the compactified variables (at the new $\tilde{X}_2$ coordinates).

The two critical points B and P$_2$ are the same as in the $\lambda_p<0$ case, being the unstable radiation dominated repeller $\Omega_r=1$ and the matter dominated saddle $\Omega_m=1$ respectively. The line L1 splits the phase space, as shown by  Figs.~\ref{fig:beltran3}--\ref{fig:beltran5}, and it is quite manifest how the trajectories change direction near this line. This is due to a sign change in the dynamical equations near L1. On the lower part of the phase space, on the other side of the line L1, we now have the new critical point P$_1$ and the asymptotic point\footnote{Note that the point $\tilde{\rm{A}}$ is not the same as the point A introduced previously. In fact, $\tilde{\rm{A}}$ is not a critical point of the system (\ref{beltran eqs1})-(\ref{beltran eqs2}), hence not appearing in Table~\ref{tab2}. But due to the nature of the physical phase space constrained by (\ref{beltran friedmann}), with all trajectories below L1 originating at this point, we have labelled it accordingly. The physical quantities $q$ and $w_{\rm{eff}}$ in Table~\ref{tab3} at point A do not apply at $\tilde{\rm{A}}$.} $\tilde{\rm{A}}$ both along $Z=0$. At point $\tilde{\rm{A}}$ one finds that $X_2 = \Omega_r \rightarrow \infty$ and $X_1 = \Omega_m \rightarrow \infty$, whilst at P$_1$ we have $X_2= \Omega_r = 0$ and $X_1 = \Omega_m \rightarrow \infty$. At P$_1$ the effective equation of state parameter is $w_{\rm{eff}}=-2$, representing a phantom regime. The new point $\tilde{\rm{A}}$ can be seen as an early time repeller as all  trajectories in the bottom section of the phase space originate there. This is best interpreted as a Big Bang like state where both matter sources diverge. This is consistent with the standard GR framework where $\rho_{m} \propto a^{-3}$ and $\rho_{r} \propto a^{-4}$ which both diverge as $a\rightarrow 0$.

Figs.~\ref{fig:beltran3} and~\ref{fig:beltran4} show the phase space for $\lambda > - \sqrt{3 \lambda_p}$, with the phase space beginning to pinch off in Fig.~\ref{fig:beltran4} as $\lambda$ decreases. The point $\tilde{\rm{A}}$ at $\tilde{X}_2=1$ corresponds to the (asymptotic) point with $X_2 \rightarrow \infty$. The points P$_2$ and B are all still present, and P$_1$ is now within the physical phase space, as mentioned above. The dashed (red) line L1 separates the phase space into two disconnected parts, with trajectories approaching the line from either side. Here the analysis is largely the same as the 1D case in the previous section, with all trajectories attracted to the singular line L1. The relative magnitudes of the parameters determine the shape of the phase space.

For the case with $\lambda < -\sqrt{3 \lambda_p}$, the phase space splits into two disjoint parts along the line L1. Fig.~\ref{fig:beltran5} shows the limiting case with $\lambda = -\sqrt{3 \lambda_p}$. The late-time de Sitter points P$_n$ exist for $\lambda \leq -\sqrt{3 \lambda_p}$ and are solutions to the equation
\begin{align}
    n(Z)= \frac{Z^2(36+12 \lambda -\lambda_p) - \lambda_p + 2 Z(\lambda_p-6\lambda)}{\lambda_p - 2 \lambda_p Z + Z^2 (36+\lambda_p)} = 2 \ ,
\end{align}
which has two solutions in the physical $Z$-range when the inequality is strict, see Fig.~\ref{fig:beltran6}. In the particular case  $\lambda = -\sqrt{3 \lambda_p}$, the two points merge into one, which coincides with the line L1 (Fig.~\ref{fig:beltran5}). The location of the `first' P$_n$ is on this singular line, whilst for smaller values of $\lambda$ the `second' point P$_n$ appears. As mentioned in the stability analysis, the points P$_n$ are always stable, and they attract all trajectories in their respective sections of the phase space. As $\lambda$ decreases relative to $\lambda_p$, the P$_n$'s move apart along the $Z$-axis. This is an interesting bifurcation rarely discussed before in cosmological models.

\section{Conclusions}

We looked at cosmological dynamical systems arising in different modified theories of gravity. By focusing on theories with second order field equations -- for example, those in $f(\mathbf{G})$, $f(T)$ and $f(Q)$ gravity -- a model independent approach could be introduced. This is accomplished by introducing the standard cosmological matter variables along with a dynamical variable directly related to the Hubble function $H$, which itself is directly related to the geometric scalars $\mathbf{G}$, $T$ and $Q$ which define the different theories. This is a unique approach that has not been studied before and is only applicable to cosmological equations that are of second order in the derivatives of the metric. For example, in $f(R)$ gravity the Ricci scalar includes terms proportional to both $H$ and $\dot{H}$, meaning the variables chosen here are insufficient to close the system. In our case, the system will always be closed once a function $f$ is specified because any function of $\mathbf{G}$, $T$ or $Q$ can be written as a function of $H$. This can also include non-minimal couplings of matter provided the equations remain second order.

The dimensionality of the dynamical systems constructed using these variables is equal to the number of matter sources, making a very general analysis of two-fluid cosmologies possible. In the past, such models required a higher dimensional approach. In fact, even without specifying a given model $f$, we have shown it is possible to determine all of the critical points of the system (Table~\ref{tab2}) as well as most of their physical properties (Table~\ref{tab3}). For example, we generally find stable points corresponding to the de Sitter solutions with effective equation of state parameter $w_{\rm{eff}}=-1$ and deceleration parameter $q=-1$. This leaves little room for models that exhibit entirely different qualitative properties than those studied here, although we note that some of the points' properties depend on the free functions $m(Z)$ and $n(Z)$ (which led to a phantom equation of state at P$_1$ in the Beltr{\'a}n model). We also note that more complicated functions $f$ will lead to a more complicated modified Friedmann constraint equation. It is the latter that determines the physical phase space for each model, and this depends on the values of all parameters in $f$. It is also this constraint equation that determines which of the critical points are physically important, and this can take a slightly convoluted form, as in one of the models we discussed.

Our model independent approach can be used to discriminate between contending models and constrain their free parameters. To discriminate between the classes of geometric theories themselves, one must look beyond the background cosmological dynamics, which are equivalent. Any viable cosmological background solution, for instance in $f(T)$ gravity, can also be realised in one of the other geometric formulations. As is clear from the approach taken here, this work is limited to considering just those background dynamics, but it would be interesting to consider how they differ beyond this level.

Despite the large freedom in the choice of functions $f$ and models that can be explored, we have shown that the qualitative dynamics remain simple enough to be captured in the general analysis performed here. With this in mind, one could look to see if observational constraints on  physical quantities could be used to constrain the theory space of all possible allowed models. At the simplest level, this would mean only considering models exhibiting a late-time de Sitter point, along with the matter saddle and an early-time radiation repeller. A more detailed treatment would be to use the effective equation of state and deceleration parameters to rule out models that are in conflict with observations, or to give tighter constraints on the free parameters in those models.

\acknowledgments
Ruth Lazkoz was supported by the Spanish Ministry of Science and Innovation through research projects FIS2017-85076-P (comprising FEDER funds), and also by the Basque Government and Generalitat Valenciana through research projects  GIC17/116-IT956-16 and  PROMETEO/2020/079 respectively.

Erik Jensko is supported by EPSRC Doctoral Training Programme (EP/R513143/1).

\appendix

\section{Beltran Model - Pathological Case} 
\label{Appendix Beltran}

For the Beltran model of Section~\ref{subsection: beltran}, the functions $m(Z)$ and $n(Z)$ share the same denominator: polynomials in $Z$ of degree 2 including the free parameter $\lambda_p$. They therefore diverge at the same $Z$ value, as a function of $\lambda_p$,
\begin{align}
  \label{Z root}
  Z= \frac{6\sqrt{-\lambda_p}+\lambda_p}{36+\lambda_p} =: Z_p \,,
\end{align}
which is valid for $\lambda_p <0$. Note that only $Z$ values in the range $[0,1]$ are physical, so the other root at $Z= (- 6\sqrt{\lambda_p}+\lambda_p)/(36+\lambda_p)$ is not considered.

The numerator of $n(Z)$, which is also a polynomial in $Z$ of degree 2, also includes the free parameter $\lambda$, which does not appear in the root of the denominators or in $m(Z)$ at all. If we expand both $m(Z)$ and $n(Z)$ infinitesimally around the $Z_p$ given by~(\ref{Z root}), one obtains at order $1/\epsilon$
\begin{align}
  m(Z) &= \frac{1}{\epsilon} \frac{12 \sqrt{-\lambda_p}}{(6+ \sqrt{-\lambda_p})^2} + \mathcal{O}(\epsilon)^0 \, , \\
  n(Z) &= \frac{1}{\epsilon} \frac{6(\sqrt{-\lambda_p}-\lambda)}{(6+\sqrt{-\lambda_p})^2 } + \mathcal{O}(\epsilon)^0  \,. \label{n root}
\end{align}
The parameter $\lambda$ can then be chosen such that the diverging term in (\ref{n root}) vanishes, $\lambda = \sqrt{-\lambda_p}$. With this choice, $n(Z)$ and $m(Z)$ can be written in the following way
\begin{align}
  m(Z) &= -\frac{4\lambda_p (Z-1)^2}{\Big( Z- Z_p \Big) \Big(6\sqrt{-\lambda_p} -\lambda_p + Z(36+\lambda_p) \Big)} \,, \\
  n(Z) &= \frac{\Big( Z- Z_p \Big) \big(12\sqrt{-\lambda_p}+ 36 -\lambda_p \big)}{6\sqrt{-\lambda_p} - \lambda_p + Z(36+ \lambda_p)} \,.
\end{align}
The term in both denominators is nonzero for all $Z \in [0,1]$, and it is clear that at $Z=Z_p$ the function $n(Z)$ is zero whilst $m(Z)$ diverges. This is the pathological case mentioned in Section~\ref{subsection: beltran}, as now there are no solutions to $n(Z)=2$. This means the point P$_n$ no longer exists. However, because $n(Z)$ now stays finite whilst $m(Z)$ diverges, the existence conditions for the critical point P$_m$ are satisfied, so P$_n$ is replaced by P$_m$. Also note that both P$_n$ and P$_m$ behave as de Sitter attractors with deceleration parameter $q=-1$ and $w_{\rm{eff}}=-1$, so the physics remains the same despite the changing of fixed points.

%\bibliographystyle{h-physrev}
%\bibliography{maindynbib}

\end{document}